# MAIN PARAMETERS OF ILC-TEVATRON BASED LEPTON-HADRON AND PHOTON-HADRON COLLIDERS

O. Çakır, A.K. Çiftçi, Ankara University, 06100 Ankara, Turkey

E. Recepoğlu, Turkish Atomic Energy Authority, 06690 Ankara, Turkey

S. Sultansoy, Gazi University, 06500 Ankara, Turkey

Ö. Yavaş, Ankara University, 06100 Ankara, Turkey

*Abstract*

The construction of the ILC tangential to Tevatron ring will give opportunity to investigate electron-proton, positron-proton, electron-antiproton, positron-antiproton interactions at 1 TeV center of mass energy. The analysis of the lepton-hadron collisions in these energy region is very important both for understanding of strong interaction dynamics and for adequate interpretation of future LHC and VLHC data. In addition, ILC-Tevatron collider will provide a possibility to realize photon-hadron collisions in the same energy region using Compton backscattered laser photon off ILC electron beam. Main parameters of these colliders are estimated and their physics search potential is briefly discussed.

## INTRODUCTION

It is known that linac-ring type machines present the most realistic way to achieve TeV scale in lepton-hadron collisions (see [1] and references therein). If the construction of ILC is realized at the FNAL site the electrons and positrons from the ILC could be collided with protons and anti-protons from the Tevatron resulting in the center of mass energy of 1 TeV lepton-hadron collisions. Using Compton backscattering of laser photons on the ILC electron beam will provide a possibility to realize photon-hadron collisions in the same energy region.

## BEAM PARAMETERS

The machine parameters and the luminosity goals of the International Linear Collider (ILC) have been discussed at the ILC workshops [2]. The most contentious choice in a parameter set will likely be the initial accelerator gradient. The gradients that have been discussed range from 18 MV/m to 45 MV/m. The specification of the peak luminosity allows the definition of a nominal parameter set which is similar to that in the TESLA TDR [3]. Table 1 lists the suggested beam parameters for ILC at 500 GeV center of mass energy. Last column of the table shows the beam parameters for Tevatron Run IIb [4].

Table 1: Main parameters of ILC and Tevatron Run IIb

| **Collider** | **ILC** | **Tevatron** | |
|---|---|---|---|
| | | **p** | **anti-p** |
| E (GeV) | 250 | 1000 | |
| T(ns) | 307.7 | 132 | |
| $f_{rep}$(Hz) | 5 | -- | |
| $\sigma_z$ (m) | $3\text{x}10^{-4}$ | 0.37 | |
| $\sigma_{x/y}(10^{-6}\text{m})$ | 0.655/0.0057 | 39 | 31 |
| $\beta_{x/y}(10^{-2}\text{m})$ | 2.1/400 | 35 | |
| $n_b$ | 2820/pulse | 103/ring | |
| $N(10^{10})$ | 2 | 27 | 10 |

## EP COLLISIONS

The achievable luminosity for electron-proton collider (ILC/e-Tevatron/p) is constrained by the electron beam power and, the intra-beam scattering which limits the emittance of the proton beam, and minimum value of the beta function which is achievable within the practical limits of focusing at the interaction point (IP). In the limit of short bunches and assuming head-on collisions, the luminosity *L* is given by

$$L = \frac{N_e N_p f_c}{2\pi\sqrt{(\sigma_{ex}^2 + \sigma_{px}^2)(\sigma_{ey}^2 + \sigma_{py}^2)}} \qquad (1)$$

where $\sigma_{e,p}$ are transverse sizes of electron and proton beams at IP, $N_e$ and $N_p$ are the numbers of electrons and protons per bunch, $f_c$ is the collision frequency.

With parameters given in Table 1, the luminosity for *ep* ($e\bar{p}$) collisions is calculated to be $8\cdot10^{29}$ cm$^{-2}$s$^{-1}$ ($4.6\cdot10^{29}$ cm$^{-2}$s$^{-1}$). The THERA [5] like upgrade of the proton beam parameters (namely, $\sigma_p$=10μm with $\beta_p$=10 cm) leads to $L_{ep} = 1.2\cdot10^{31}$ cm$^{-2}$s$^{-1}$. In addition the luminosity can be increased by an extra factor ~3 by applying a "dynamic focusing" scheme, where the p-beam waist travels with the e-bunch during collision [6].

The physics search potential of the ep option of the ILC⊗Tevatron collider is the same as THERA. A detailed study can be found in the THERA Book [5]. The partial list of physics goals of the ILC⊗Tevatron based ep collider includes:

- The extension of the kinematic range down to $x \cong 10^{-6}$ allows access to the high parton density domain and its detailed exploration in the deep inelastic regime.

- The study of forward going jets at ILC⊗Tevatron is expected to reveal the mechanism for the evolution of QCD radiation at low x. The increased range for the $Q^2$ evolution of parton densities will allow a precision measurement of $\alpha_s$.

- ILC⊗Tevatron will probe physics beyond the Standard Model. In particular, leptoquarks or squarks in supersymmetry with R-parity violation can be produced and their couplings determined in a rather complete manner. ILC⊗Tevatron is very sensitive to four-fermion contact interactions and probes compactification scales up to about 2.8 TeV via t-channel exchange of Kaluza-Klein gravitons in models with large extra dimensions. ILC⊗Tevatron will extend the searches for excited fermions to masses of up to 1 TeV.

- The photon structure will be resolved at harder scales and lower $x_\gamma$. The higher cross section for heavy-flavour production will permit the charm and bottom content of the quasi-real and virtual photon to be explored.

- Polarized electron-proton scattering at ILC⊗Tevatron allows the study of the spin structure of the proton and its theoretical interpretation in QCD to be extended into an hitherto unexplored kinematic range of low x and large $Q^2$.

Additional lepton-anti-proton collider options could be an advantage of the ILC⊗Tevatron.

## γP COLLISIONS

The idea of constructing γp colliders on the base of linac-ring type ep machines was considered in [7]. The details of the method used to obtain high energy photon beam from Compton backscattering of laser light off a beam of high energy electrons can be found in [8], which was considered earlier for γe and γγ colliders.

The luminosity of the γp collider depends on the distance z between conversion point and collision point [9], where some design problems were considered. Taking into account the photon/electron conversion ratio $N_\gamma/N=0.65$ we obtain a luminosity $L_{\gamma p}= 5.2\cdot10^{29}$ cm$^{-2}$ s$^{-1}$ at z=0 for ILC⊗Tevatron with 250 GeV electron beams (with THERA like upgrade $L_{\gamma p}= 8\cdot10^{30}$ cm$^{-2}$ s$^{-1}$). The luminosity decreases slightly with increasing values of the parameter z (a factor of 0.5 at z = 5 m) and opposite helicity values for laser and electron beams are advantageous (see Fig. 1 and Fig. 2). If one assumes head-on collisions and exclude deflection of electrons (to avoid synchrotron radiation and passage from focusing quadrupoles), the residual electron beam will collide with proton beam together with high energy photon beam, but because of larger cross sections of γp collisions the background resulting from ep collisions may be neglected. With a few (~4) micro radians scattering angle the residual electron beam transverse size will be 100 μm at the distance of 25 m from conversion region and the focusing quadrupoles for proton beam have negligible influence on the residual electrons.

The physics search potential of γp colliders is reviewed in [10]. The partial list of physics goals of the ILC⊗Tevatron based γp collider includes:

- The total cross section at TeV scale will be $\sigma(\gamma p \to \text{hadrons}) \approx 100\ \mu b$
- Double-jets events with $p_T$>100 GeV are expected to be $10^4$ per one working year
- Heavy quark pairs, $10^7$, $10^6$ and $10^2$ events per one working year for $c\bar{c}$, $b\bar{b}$ and $t\bar{t}$, respectively
- Single W and Z production with anomalous interactions (~$10^4$ events/year)
- Anomalous production of t-quark and the fourth SM family quarks via γq fusion
- Hadronic structure of photon can be examined at γp option of the ILC⊗Tevatron collider
- Excited quarks u* or d* with m ≤1 TeV.

This option will give unique opportunity to investigate small $x_g$ region via $c\bar{c}$ and $b\bar{b}$ pair production [10]. Due to the advantage of the real γ spectrum heavy quarks will be produced via γg fusion at characteristic

$$x_g \approx \frac{5.6\times m_{c(b)}^2}{0.83\times s_{ep}} \qquad (2)$$

which is approximately $10^{-5}$ ($10^{-4}$) for charmed (beauty) hadrons.

## CONCLUSION

The physics phenomena at extremely small x l sufficiently high $Q^2$ is very important for understandi the nature of strong interactions at all levels fr nucleons to partons. ILC-Tevatron collider will provid possibility to realize TeV energy range lepton-hadron a photon-hadron collisions in the same energy region usi Compton backscattered laser photon off ILC electr beam.

## ACKNOWLEDGEMENT

This work is supported by Turkish Atomic Ener Authority and Turkish State Planning Organization under the Grants No DPT-2002K-120250, DPT-2003K-1201906-5.

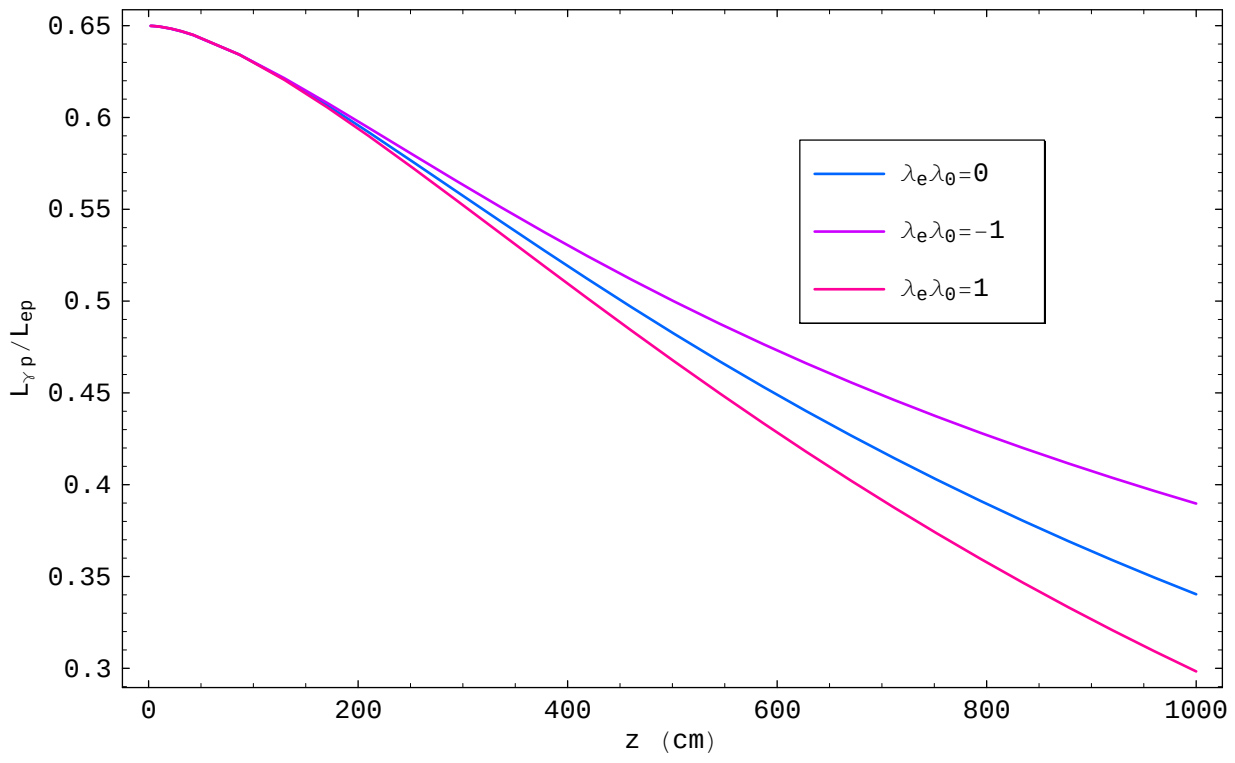

Fig. 1. The dependence of luminosity on the distance z for different beam polarization at γp collider.

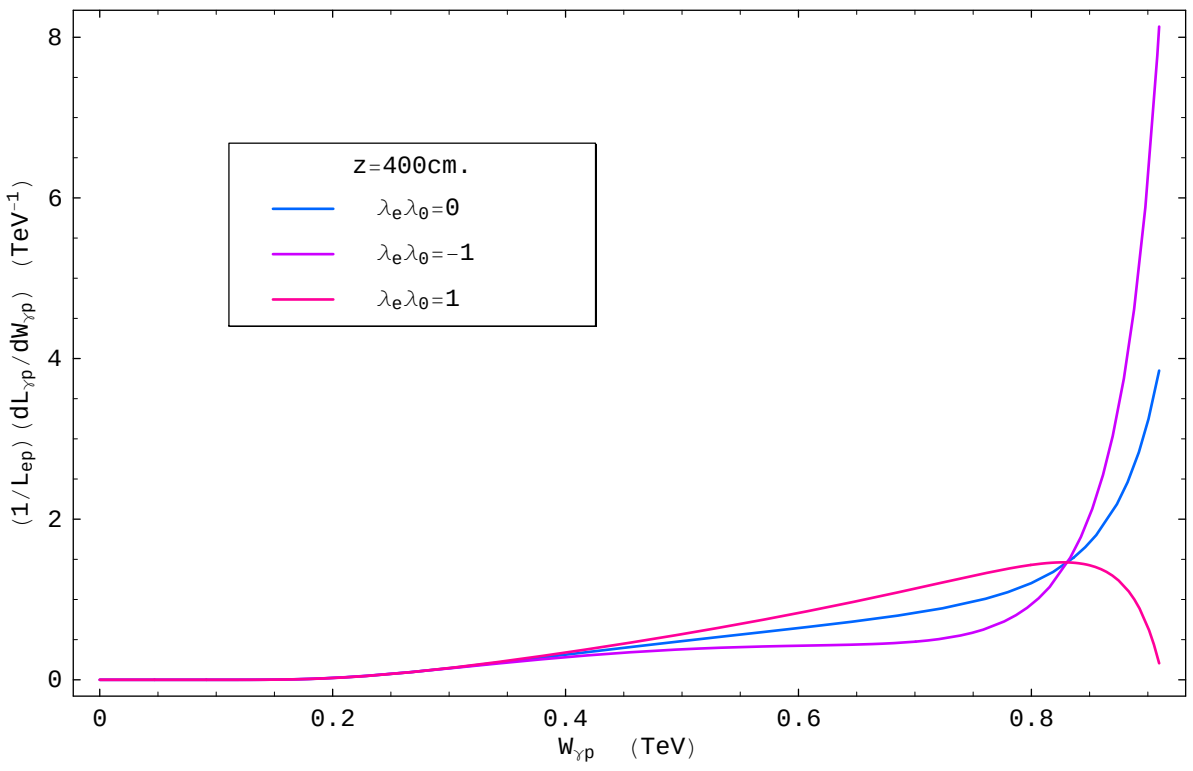

Fig. 2. Luminosity distribution as a function of γp invariant mass.